\documentclass[aps,nofootinbib,longbibliography,prd,onecolumn]{revtex4}
\usepackage[a4paper,top=2cm,bottom=2cm,left=2cm,right=2cm]{geometry}
\usepackage[colorlinks,citecolor=blue,linkcolor=blue,urlcolor=blue]{hyperref}
\usepackage[babel]{csquotes}
\usepackage{graphicx}
\usepackage{xcolor}
\usepackage{amsmath,amssymb,amsfonts}
\usepackage[applemac]{inputenc}
\usepackage{mathtools,mathrsfs}
\usepackage{enumerate}
\usepackage{dsfont}
\usepackage{array}
\usepackage{tabularx}
\usepackage{scrextend}
\usepackage{float}
\usepackage{comment}
\usepackage[default]{gillius}
\usepackage{lipsum}
\usepackage[
 format = hang, 
 justification = RaggedRight, 
 textfont = it, 
 figurename = Figure,
 ]{caption}
 
\newcommand{\dd}{{\rm d}}
\newcommand{\rn}[1]{\textcolor{red}{#1}}
\newcommand{\transpose}{\intercal}

\begin{document}

\title{Classification of Primary Constraints of Quadratic Non-Metricity Theories of Gravity}
\author{Fabio D'Ambrosio}\email{fabioda@phys.ethz.ch}
\author{Lavinia Heisenberg} \email{lavinia.heisenberg@phys.ethz.ch}
\affiliation{Institute for Theoretical Physics,
ETH Zurich, Wolfgang-Pauli-Strasse 27, 8093, Zurich, Switzerland}

\begin{abstract}
\noindent We perform the ADM decomposition of a five-parameter family of quadratic non-metricity theories and study their conjugate momenta. After systematically identifying all possible conditions which can be imposed on the parameters such that different sets of primary constraints arise, we find that the five-parametric theory space can be compartmentalized into nine different sectors, based on the presence or absence of primary constraints. This classification allows to dismiss certain classes of theories as unphysical and invites further investigations into the remaining sectors, which may contain phenomenologically interesting modifications of General Relativity. 
\end{abstract}

\maketitle

\section{Introduction}
Gravity and geometry have been linked together ever since Einstein's brilliant formulation of General Relativity (GR), which conceptualizes spacetime as a differential manifold endowed with a metric tensor and ascribes the effects of gravitation to the curvature of spacetime. At the base of this interpretation of gravitation rests the equivalence principle, but also two seemingly innocuous mathematical postulates. In fact, a manifold cannot only be endowed with a metric, but also with an affine structure described by a connection $\Gamma^{\alpha}_{\ \mu\nu}$. In the standard formulation of GR, the connection is postulated to be torsion-free and metric-compatible and these two requirements uniquely determine $\Gamma^{\alpha}_{\ \mu\nu}$ to be the Levi-Civita connection induced by the metric tensor.

In general, however, the metric and the connection are independent objects and two alternative but equivalent geometric descriptions of gravity are conceivable once the requirements of torsion-freeness and metric-compatibility are relaxed: The Teleparallel Equivalent formulation of GR (TEGR) \cite{TeleparallelBook,Cai:2015,Krssak:2018} is based on a flat and metric-compatible connection, ascribing the effects of gravity to torsion, while the Symmetric Teleparallel Equivalent of GR (STEGR) \cite{Jimenez:2018} builds on a flat, torsionless connection and describes gravity via the non-metricity tensor.

Despite their equivalence to GR, the teleparallel formulations are appealing from a theoretical perspective as becomes apparent when the actions of the various theories are considered. Recall that the Einstein-Hilbert action leads to an ill-posed variational problem due to the presence of second order derivatives in the Riemann curvature tensor. This requires to not only impose boundary conditions on the metric, but also on its normal derivatives. Alternatively, one can achieve a cancellation of the higher derivative terms by introducing the Gibbons-Hawking-York boundary term. Teleparallel theories, however, do not suffer from this problem as their actions only contain first order derivatives. This property of teleparallel theories naturally leads to second order field equations, without the need for alterations to the action principle.

Symmetric teleparallel theories possess one additional property which makes them attractive: The vanishing of curvature and torsion implies the existence of a gauge -- the so-called coincident gauge~\cite{BeltranJimenez:2017} -- which globally trivializes the connection, $\Gamma^\alpha_{\ \mu\nu} = 0$. This makes STEGR interesting for applications in numerical relativity and it might shed new light on issues of canonical quantum gravity. Moreover, extensions of STEGR could have implications for inflation and dark energy~\cite{Jimenez:2019} as well as dark matter phenomenology. 

However, before studying phenomenological models based on symmetric teleparallelism, it is essential to identify those theories which are self-consistent. This involves counting the degrees of freedom which are being propagated and determining whether these candidate theories harbor ghosts. These questions can be addressed systematically within the generalized Hamiltonian formalism pioneered by Dirac and Bergmann~\cite{Dirac:1950,Bergmann:1951,DiracBook}, of which we will make extensive use in this article. Our starting point in section~\ref{sec:SymmetricTeleparallelism} is an action functional constructed from arbitrary linear combinations of the five independent non-metricity scalars. We then proceed in section~\ref{sec:ADM} by performing the $3+1$ decomposition of this action in terms of ADM variables, thereby always working in the coincident gauge where the connection is trivial. In order to transition to the Hamiltonian formalism, we compute all conjugate momentum densities of the ADM fields and find, unsurprisingly, that the number of primary constraints in the Hamiltonian theory depends on how the five parameters of the action are chosen.

In section~\ref{sec:Classification} we systematically analyze all possible conditions that can be imposed on the five action parameters. This leads to a compartmentalization of the five-parametric theory space into nine distinct sectors, where each sector is characterized by a certain number of primary constraints. This classification of primary sectors can already be applied to dismiss certain theories as unphysical -- either because there are too many or too few primary constraints. The surviving sectors invite further investigations along the lines of a full-fledged Hamiltonian analysis.

\section{Symmetric Teleparallelism}\label{sec:SymmetricTeleparallelism}
Let us start by introducing a non-Riemannian, affine manifold described by $(\mathcal M, g_{\mu\nu}, \Gamma^{\alpha}_{\ \mu\nu})$, where $g_{\mu\nu}$ denotes the components of the metric tensor of signature $(-,+,+,+)$ and $\Gamma^{\alpha}_{\ \mu\nu}$ represents an affine connection. This connection defines a notion of covariant differentiation through its action on vectors and co-vectors,
\begin{align}
	\nabla_\mu V^\alpha &= \partial_\mu V^\alpha + \Gamma^\alpha_{\ \mu\lambda} V^\lambda\notag\\
	\nabla_\mu V_\alpha &= \partial_\mu V_\alpha - \Gamma^\lambda_{\ \mu\alpha} V_\lambda,
\end{align}

and it can be used to describe three independent geometric properties of a spacetime: curvature, torsion, and non-metricity. The first two objects, curvature and torsion, are defined by
\begin{align}\label{eq:SymParaCond}
	R^\alpha_{\ \beta\mu\nu} &:= 2 \partial_{[\mu}\Gamma^\alpha_{\ \nu]\beta} + 2 \Gamma^\alpha_{[\mu|\lambda|}\Gamma^\lambda_{\ \nu]\beta}\notag\\
	T^\alpha_{\ \mu\nu} &:= 2\Gamma^\alpha_{\ [\mu\nu]},
\end{align}
and symmetric teleparallelism demands that both tensors vanish. The vanishing of the curvature tensor implies that the connection must have the form
\begin{equation}
	\Gamma^\alpha_{\ \mu\nu} = \left(\Lambda^{-1}\right)^\alpha_{\ \rho} \partial_\mu \Lambda^\rho_{\ \nu},
\end{equation}
where $\Lambda^\alpha_{\ \beta}\in GL(4,\mathbb R)$. The requirement of vanishing torsion further restricts the matrix $\Lambda^\alpha_{\ \beta}$ to have the form $\Lambda^\alpha_{\ \beta} = \partial_\beta \xi^\alpha$, for arbitrary $\xi^\alpha$, and the connection consequently becomes
\begin{equation}\label{eq:CoincidentConnection}
	\Gamma^\alpha_{\ \mu\nu} = \frac{\partial x^\alpha}{\partial \xi^\lambda}\partial_\mu\partial_\nu\xi^\lambda.
\end{equation}
What is remarkable about~\eqref{eq:CoincidentConnection} is that the connection can be set globally\footnote{There might be topological obstructions, but such considerations are outside the scope of the current paper.} to zero by the affine gauge choice $\xi^\alpha = M^{\alpha}_{\ \beta} x^\beta + \xi^\alpha_0$, where $M^{\alpha}_{\ \beta}$ is a non-degenerate matrix with constant entries and $\xi^\alpha_0$ is a constant vector. This is known as the coincident gauge~\cite{BeltranJimenez:2017}. With curvature and torsion set to zero, the non-metricity tensor is the only remaining non-trivial object. As it measures the failure of the connection~\eqref{eq:CoincidentConnection} to be metric-compatible, it is defined by
\begin{equation}
	Q_{\alpha\mu\nu} := \nabla_\alpha g_{\mu\nu} = \partial_\alpha g_{\mu\nu} - 2 \frac{\partial x^\rho}{\partial\xi^\lambda}\partial_\alpha\partial_{(\mu}\xi^\lambda g_{\nu)\rho}.
\end{equation}
At quadratic order, there are only five independent scalars that can be built from the non-metricity tensor,
\begin{equation}\label{eq:GenNonMetricityScalar}
	\mathbb Q := c_1\, Q_{\alpha\beta\gamma}Q^{\alpha\beta\gamma} + c_2\, Q_{\alpha\beta\gamma}Q^{\beta\alpha\gamma}	 + c_3\, Q_\alpha Q^{\alpha} + c_4\, \bar{Q}_\alpha\bar{Q}^\alpha + c_5\, Q_\alpha \bar{Q}^\alpha,
\end{equation}
where $c_i$ are arbitrary real numbers and $Q_\alpha:= Q_{\alpha\nu}{}^{\nu}$ and $\bar{Q}_\alpha:= Q^\nu{}_{\nu\alpha}$ denote the two independent traces. In terms of the non-metricity scalar $\mathbb Q$, the covariant five-parametric action of symmetric teleparallelism can be written as
\begin{equation}\label{eq:CovQuadLagrangian}
	\mathcal S[g, \Gamma; \lambda, \rho] := \int_\mathcal{M}\dd^4 x\,\left(\frac{1}{2}\sqrt{-g}\,\mathbb Q + \lambda_\alpha{}^{\beta\mu\nu} R^\alpha{}_{\beta\mu\nu} + \rho_\alpha{}^{\mu\nu} T^\alpha{}_{\mu\nu}\right),
\end{equation}
where the tensor densities $\lambda_\alpha{}^{\beta\mu\nu}$ and $\rho_\alpha{}^{\mu\nu}$ act as Lagrange multipliers which force the spacetime to be flat and torsionless. If the $c_i$ parameters are chosen as
\begin{equation}\label{eq:GRCoefficients}
	c_1 = -\frac14,\quad c_2 = \frac12,\quad c_3 = \frac14, \quad c_4 = 0,\quad c_5 = -\frac12,
\end{equation}
the Symmetric Teleparallel Equivalent of GR is recovered. Our objective of performing the ADM decomposition of STG and classifying all primary constraints can in principle be achieved using the action~\eqref{eq:CovQuadLagrangian}, which is a functional of ten metric components, $64$ connection components and which also depends on $44$ Lagrange multipliers. However, this task can be drastically simplified by exclusively working in the coincident gauge. This is tantamount to strongly imposing the constraints~\eqref{eq:SymParaCond} and setting the connection to zero globally. Hence, we are left with the simpler action
\begin{equation}\label{eq:CGSTG}
	\mathcal S[g] = \frac{1}{2}\int_\mathcal{M}\dd^4 x\,\sqrt{-g}\,\mathcal Q,
\end{equation}
which is solely a functional of the metric and where the scalar $\mathcal Q := \left.\mathbb Q\right|_{\Gamma = 0}$ is explicitly given by
\begin{equation}
	\mathcal Q = \left(c_1 g^{\alpha\kappa} g^{\beta\mu} g^{\gamma\nu} + c_2 g^{\alpha\nu} g^{\beta\mu} g^{\gamma\kappa} + c_3 g^{\alpha\kappa} g^{\beta\gamma} g^{\mu\nu} + c_4 g^{\alpha \beta} g^{\gamma\nu} g^{\kappa \mu} + c_5 g^{\alpha\beta} g^{\gamma\kappa} g^{\mu\nu}\right)\partial_{\alpha} g_{\beta\gamma} \partial_{\kappa} g_{\mu\nu}.
\end{equation}  
The action~\eqref{eq:CGSTG} is the starting point for the ADM decomposition and computation of conjugate momenta, which is the first step toward counting degrees of freedom and assessing the health status of the five-parameter family of theories described by $\mathcal Q$.

\section{ADM Decomposition and Conjugate Momenta}\label{sec:ADM}
In order to perform the Hamiltonian analysis of the action~\eqref{eq:CGSTG}, we introduce local coordinates $(t, x^{a})$ with $a\in\{1,2,3\}$ and foliate the manifold $\mathcal M$ into spacelike hypersurfaces $\Sigma_t$ of constant coordinate time $t$. Furthermore, it is convenient to perform an ADM decomposition where the metric degrees of freedom are given by a lapse function $N$, a shift vector field $N^{a}$, and the intrinsic three-dimensional metric $h_{ab}$ of the spacelike leaves. Concretely, the metric and its inverse are given by
\begin{equation}\label{eq:ADMmetric}
	g_{\mu\nu} = \begin{pmatrix}
		-N^2 + N_{a} N^{a} & N_{a}\\
		N_{a} & h_{ab}
	\end{pmatrix}\quad\text{and}\quad
	g^{\mu\nu} = \begin{pmatrix}
		-\frac{1}{N^2} & \frac{N^{a}}{N^2}\\
		\frac{N^{a}}{N^2} & h^{ab} - \frac{N^{a} N^{b}}{N^2}
	\end{pmatrix}.
\end{equation}
Our convention is to use lower case roman letters to denote spatial indices while the temporal index will be denoted by $\rn{0}$ in the sequel. As usual, spatial indices are raised and lowered with the intrinsic metric $h_{ab}$. Using~\eqref{eq:ADMmetric} one can easily show that the determinant of the spacetime metric is given by $\sqrt{-g} = N\sqrt{h}$. 

With these prearrangements, it is in principle a straightforward task to perform the ADM decomposition of the action~\eqref{eq:CGSTG}. It is just a matter of inserting~\eqref{eq:ADMmetric} into~\eqref{eq:CGSTG} and computing all partial derivatives and metric contractions. In practice, however, it is a tedious and unenlightening exercise. We have therefore written a code, based on the Mathematica extension package xAct~\cite{xAct}, to accomplish the ADM decomposition and perform parts of the Hamiltonian analysis.
The same file has also been used to compute the momentum densities\footnote{We use Ashtekar's tilde notation to denote tensor densities of weight one.} conjugate to lapse, shift, and intrinsic metric. These are explicitly given by
\begin{align}\label{eq:GeneralMomenta}
	\tilde\pi &:= \frac{\delta\mathcal S}{\delta\dot N} = \frac{\sqrt{h}}{N^4}\bigg[2\tilde c \left( Q_{\rn{000}} - N^{a} Q_{a\rn{00}} - 2N^{a} Q_{\rn{00}a} + 2 N^{a} N^{b} Q_{ab\rn{0}} + N^{a} N^{b} Q_{\rn{0} ab} - N^{a} N^{b} N^{c} Q_{abc}\right) \notag\\
	 &\phantom{:= \frac{\delta\mathcal S}{\delta N} = \frac{\sqrt{h}}{N^4}(2} + c_{35} \left(N^2 N^{a} Q_a - N^2 Q_{\rn{0}}\right) + c_{45}\left(N^2 N^{a} \bar Q_a - N^2  \bar Q_{\rn{0}}\right)\bigg]\notag\\
	 \tilde\pi_a &:= \frac{\delta \mathcal S}{\delta\dot N^{a}} = \frac{\sqrt{h}}{2N^3}\bigg[2\hat c\left(Q_{\rn{00}a}-N^{b}Q_{ba\rn{0}} - N^{b} Q_{\rn{0} ab} + N^{b} N^{c} Q_{bac}\right) + c_{25}\left(Q_{a\rn{00}} -2 N^{b} Q_{ab\rn{0}} + N^{b} N^{c} Q_{abc}\right)\notag\\
	 &\phantom{:= \frac{\delta \mathcal S}{\delta\dot N^{a}} = \frac{\sqrt{h}}{2N^3}[2} - 2 c_4 N^2 \bar Q_a - c_5 N^2 Q_a\bigg]\notag\\
	 \tilde\pi^{ab} &:= \frac{\delta\mathcal S}{\delta\dot h_{ab}} = \frac{\sqrt{h}}{2 N^3}\bigg[ c_{35} \left(h^{ab} Q_{\rn{000}} - h^{ab} N^{c} \left(Q_{c\rn{00}} + 2 Q_{\rn{00} c} - N^{d} \left\{ 2 Q_{cd\rn{0}} + Q_{\rn{0} cd} - N^{e}Q_{cde}\right\}\right)\right)\notag\\
	 &\phantom{:= \frac{\delta\mathcal S}{\delta\dot h_{ab}} = \frac{\sqrt{h}}{2 N^3}[c} + N^2 N^{c} h^{ad} h^{be}  \left(2 c_1 Q_{cde} + c_2 Q_{dce} + c_2 Q_{ecd}\right) + N^2 N^{c} h^{ab} h^{de} \left(2c_3 Q_{cde} + c_5 Q_{dce}\right)\notag\\
	&\phantom{:= \frac{\delta\mathcal S}{\delta\dot h_{ab}} = \frac{\sqrt{h}}{2 N^3}[c} - N^2 h^{ab} h^{cd}\left(2 c_3 Q_{\rn{0}cd} + c_5 Q_{cd\rn{0}}\right) - N^2 h^{ac} h^{bd} \left(2 c_1 Q_{\rn{0} cd} + c_2 Q_{cd\rn{0}} + c_2 Q_{dc\rn{0}}\right)\bigg],
\end{align} 
where we introduced the shorthand notations
\begin{align}\label{eq:ShortHandNotations}
	\tilde c &:= c_1 + c_2 + c_3 + c_4 + c_5 \notag\\
	\hat c &:= 2c_1 + c_2 + c_4 \notag\\
	c_{i5} &:= 2c_i + c_5 \quad\text{for } i\in\{2,3,4\}
\end{align}
and we kept the non-metricity components for compactness of notation. We find that in general the momenta conjugate to lapse and shift do not vanish. Even for the parameter choice~\eqref{eq:GRCoefficients}, which corresponds to Coincident GR, one finds the non-vanishing momentum densities
\begin{align}\label{eq:GRMomenta}
	\left.\tilde\pi\right|_{c_i\to\text{GR}} &= \frac{\sqrt{h}}{2 N^2}\left(\bar Q_{\rn{0}} - N^{a} \bar Q_{a}\right)\notag\\
	\left.\tilde\pi_a\right|_{c_i\to\text{GR}} &= \frac{\sqrt{h}}{4 N^3}\left(Q_{a\rn{00}} + N^2 h^{bc}Q_{abc} + N^{b}(Q_{abc} N^{c} - 2 Q_{ab\rn{0}})\right)\notag\\
	\left.\tilde\pi^{ab}\right|_{c_i\to\text{GR}} &= \frac{\sqrt{h}}{4N} \bigg[N^{c} h^{ad} h^{be}  \left(Q_{dce} + Q_{ecd} - Q_{cde}\right) + N^{c} h^{ab} h^{de} \left(Q_{cde} - Q_{dce}\right)\notag\\
	&\phantom{\frac{\sqrt{h}}{4N} \bigg[N^2 } - h^{ab} h^{cd}\left(Q_{\rn{0}cd} - Q_{cd\rn{0}}\right) - h^{ac} h^{bd} \left(Q_{cd\rn{0}} + Q_{dc\rn{0}} - Q_{\rn{0} cd}\right)\bigg].
\end{align}
This is consistent with the observation made in~\cite{DAmbrosio:2020} where a detailed Hamiltonian analysis for Coincident General Relativity was carried out after the momenta were made to vanish through the addition of appropriate boundary terms.  Also, notice that in~\eqref{eq:GRMomenta} the momenta conjugate to lapse and shift do not contain any velocity fields. Time derivatives only appear in non-metricity components of the form $Q_{\rn{0}\mu\nu}$ and in the trace $Q_{\rn{0}}$ (but not in $\bar Q_{\rn{0}}$). It therefore follows that the momenta conjugate to lapse and shift constitute primary constraints, despite being different from zero, while the momenta conjugate to the intrinsic metric depend on $Q_{\rn{0}cd} = \dot h_{cd}$ and are therefore dynamical. 

This observation is immediately transferable to the general case described by~\eqref{eq:GeneralMomenta}: The non-vanishing momenta can be turned into primary constraints by imposing appropriate conditions on the $c_i$ parameters. For instance, the choices $\tilde c = 0, c_{35} = 0$ remove all velocity terms from the momentum density~$\tilde\pi$ while the condition $\hat c = 0$ achieves the same for $\tilde\pi_a$. More elaborate parameter choices exist besides these two obvious examples and in the next section we provide a systematic derivation of all possible conditions that can be imposed on the $c_i$ parameters such that different sets of primary constraints emerge.

\section{Primary Constraints and Classification of Theories}\label{sec:Classification}
Primary constraints arise whenever it is not possible to express all velocity fields in terms of momenta. Since~\eqref{eq:GeneralMomenta} forms an inhomogeneous, linear system of equations for the velocities $(\dot N, \dot N^{a}, \dot h_{ab})$, this is tantamount to stating that constraints arise whenever the matrix describing the linear system~\eqref{eq:GeneralMomenta} is not invertible. A further moment of reflection will show that the invertibility of this matrix is equivalent to the invertibility of the Hessian
\begin{equation}\label{eq:DefHessian}
	H := -\int_{\Sigma_t} \frac{1}{\sqrt{h}}\frac{\delta^2\mathcal S}{\delta\dot{\Psi}^{I}\delta\dot{\Psi}^{J}}\,\dd^3 x,
\end{equation}
where $\dot{\Psi}^{I}$ stands representatively for $(\dot N, \dot N^{a}, \dot h_{ab})$. Its components $H_{\dot{\Psi}^{I} \dot{\Psi}^{J}}$ can be inferred from~\eqref{eq:GeneralMomenta} and are formally given by
\begin{align}\label{eq:HessianMatrixFormal}
	H_{\dot N\dot N} &= \frac{4}{N^3}\tilde c & H_{\dot N^{a} \dot N^{b}} &= -\frac{\hat c}{N^3}\, h_{ab} \notag\\
	H_{\dot N \dot N^{a}} &= 0 & H_{\dot N^{a} \dot h_{bc}} &= 0 \notag\\
	H_{\dot N \dot h_{ab}} &= \frac{c_{35}}{N^2}\,h^{ab} & H_{\dot h_{ab} \dot h_{cd}} &= \frac{1}{N} \left(c_1 h^{ca}h^{db} + c_3 h^{ab} h^{cd}\right).
\end{align}
If the rank of the Hessian is less than ten, then there are $M:=10 - \text{rank} H$ independent constraint equations among the configuration space variables. Our task is therefore to find all conditions on the $c_i$ parameters which lead to different numbers $M$ of primary constraints. To that end, it is useful to pick a specific ordering for the velocity fields and write out the Hessian matrix explicitly. Our convention is $\dot\Psi^{I} = (\dot N, \dot N^{\rn{1}}, \dot N^{\rn{2}}, \dot N^{\rn{3}}, \dot h_{\rn{11}}, \dot h_{\rn{22}}, \dot h_{\rn{33}}, \dot h_{\rn{12}}, \dot h_{\rn{13}}, \dot h_{\rn{23}})$ and this results in the matrix
\begin{equation}\label{eq:HessianExplicit}
	\scalebox{0.8}{$\begin{pmatrix}
		\frac{4}{N^3}\tilde c & 0 & 0 & 0 &  c_{35}\frac{h^{\rn{11}}}{N^2} &  c_{35}\frac{h^{\rn{22}}}{N^2} &  c_{35}\frac{h^{\rn{33}}}{N^2} &  c_{35}\frac{h^{\rn{12}}}{N^2} &  c_{35}\frac{h^{\rn{13}}}{N^2} &  c_{35}\frac{h^{\rn{23}}}{N^2} \\
		0 & -\hat c \frac{h_{\rn{11}}}{N^3} & -\hat c \frac{h_{\rn{12}}}{N^3} & -\hat c \frac{h_{\rn{13}}}{N^3} & 0 & 0 & 0 & 0 & 0 & 0\\
		0 & -\hat c \frac{h_{\rn{12}}}{N^3} & -\hat c \frac{h_{\rn{22}}}{N^3} & -\hat c \frac{h_{\rn{23}}}{N^3} & 0 & 0 & 0 & 0 & 0 & 0\\
		0 & -\hat c \frac{h_{\rn{13}}}{N^3} & -\hat c \frac{h_{\rn{23}}}{N^3} & -\hat c \frac{h_{\rn{33}}}{N^3} & 0 & 0 & 0 & 0 & 0 & 0\\
		 c_{35}\frac{h^{\rn{11}}}{N^2} & 0 & 0 & 0 & \frac{c_{13}(h^{\rn{11}})^2}{N} &  \frac{c_1 (h^{\rn{12}})^2+c_3 h^{\rn{11}} h^{\rn{22}}}{N} & \frac{c_1 (h^{\rn{13}})^2+c_3 h^{\rn{11}} h^{\rn{33}}}{N} & \frac{c_{13} h^{\rn{11}}h^{\rn{12}}}{N} & \frac{c_{13} h^{\rn{11}}h^{\rn{13}}}{N} & \frac{c_1 h^{\rn{12}} h^{\rn{13}} + c_3 h^{\rn{11}} h^{\rn{23}}}{N}\\
		 c_{35}\frac{h^{\rn{22}}}{N^2} & 0 & 0 & 0 & \frac{c_1 (h^{\rn{12}})^2+c_3 h^{\rn{11}} h^{\rn{22}}}{N} & \frac{c_{13}(h^{\rn{22}})^2}{N} & \frac{c_1 (h^{\rn{23}})^2 + c_3 h^{\rn{22}} h^{\rn{33}}}{N} & \frac{c_{13} h^{\rn{12}} h^{\rn{22}}}{N} & \frac{c_1 h^{\rn{12}} h^{\rn{23}} + c_3 h^{\rn{12}} h^{\rn{22}}}{N} & \frac{c_{13}h^{\rn{22}} h^{\rn{23}}}{N}\\
		 c_{35}\frac{h^{\rn{33}}}{N^2} & 0 & 0 & 0 & \frac{c_1 (h^{\rn{13}})^2+c_3 h^{\rn{11}} h^{\rn{33}}}{N} & \frac{c_1 (h^{\rn{23}})^2 + c_3 h^{\rn{22}} h^{\rn{33}}}{N} & \frac{c_{13} (h^{\rn{33}})^2}{N} & \frac{c_1 h^{\rn{13}} h^{\rn{23}} + c_3 h^{\rn{12}} h^{\rn{33}}}{N} & \frac{c_{13} h^{\rn{13}} h^{\rn{33}}}{N} & \frac{c_{13} h^{\rn{23}} h^{\rn{33}}}{N} \\
		 c_{35}\frac{h^{\rn{12}}}{N^2} & 0 & 0 & 0 & \frac{c_{13} h^{\rn{11}}h^{\rn{12}}}{N} & \frac{c_{13} h^{\rn{12}} h^{\rn{22}}}{N} & \frac{c_1 h^{\rn{13}} h^{\rn{23}} + c_3 h^{\rn{12}} h^{\rn{33}}}{N} & \frac{c_1 h^{\rn{11}} h^{\rn{22}} + c_3 (h^{\rn{12}})^2}{N} & \frac{c_1 h^{\rn{11}} h^{\rn{23}} + c_3 h^{\rn{12}} h^{\rn{13}}}{N} & \frac{c_{13} h^{\rn{12}} h^{\rn{23}}}{N} \\
		 c_{35}\frac{h^{\rn{13}}}{N^2} & 0 & 0 & 0 & \frac{c_{13} h^{\rn{11}}h^{\rn{13}}}{N} & \frac{c_1 h^{\rn{12}} h^{\rn{23}} + c_3 h^{\rn{12}} h^{\rn{22}}}{N}  & \frac{c_{13} h^{\rn{13}} h^{\rn{33}}}{N} & \frac{c_1 h^{\rn{11}} h^{\rn{23}} + c_3 h^{\rn{12}} h^{\rn{13}}}{N} & \frac{c_1 h^{\rn{11}} h^{\rn{33}} + c_3 (h^{\rn{13}})^2}{N} & \frac{c_1 h^{\rn{12}} h^{\rn{33}} + c_3 h^{\rn{13}} h^{\rn{23}}}{N} \\
		 c_{35}\frac{h^{\rn{23}}}{N^2} & 0 & 0 & 0 & \frac{c_1 h^{\rn{12}} h^{\rn{13}} + c_3 h^{\rn{11}} h^{\rn{23}}}{N} & \frac{c_{13}h^{\rn{22}} h^{\rn{23}}}{N} & \frac{c_{13} h^{\rn{23}} h^{\rn{33}}}{N} & \frac{c_{13} h^{\rn{12}} h^{\rn{23}}}{N} & \frac{c_1 h^{\rn{12}} h^{\rn{33}} + c_3 h^{\rn{13}} h^{\rn{23}}}{N} & \frac{c_3 (h^{\rn{23}})^2 + c_1 h^{\rn{22}} h^{\rn{33}}}{N}
	\end{pmatrix}$}
\end{equation}
where $c_{13}$ is a shorthand notation for $c_1 + c_3$. A convenient way to characterize the degeneracy of the above matrix is by computing its determinant. Naturally, this can be achieved using computer algebra systems, but it is also worth noting that this matrix consists of nested block matrices and we can therefore make deductions about its structure. In fact, in a first step we can think of $H$ as being given by
\begin{equation}
	H = \begin{pmatrix}
		\frac{4}{N^3}\tilde c & v^\transpose\\
		v & \mathbb M
	\end{pmatrix},
\end{equation} 
where the vector $v$ stands for the first column of~\eqref{eq:HessianExplicit}, not including the element $\frac{4}{N^3}\tilde c$, $v^\transpose$ stands for its transpose, and $\mathbb M$ represents the remaining $9\times 9$ block matrix which makes up~\eqref{eq:HessianExplicit}. The determinant of block matrices is easy to compute and one readily finds
\begin{equation}
	\det H = \left(\frac{4}{N^3}\tilde c - v^\transpose \mathbb M^{-1} v\right)\det\mathbb M,
\end{equation}
assuming $\mathbb M$ is invertible. Next, notice that $\mathbb M$ itself is also a block matrix which can be written as
\begin{equation}
	\mathbb M = \begin{pmatrix}
		-\frac{\hat c}{N^3} h & 0_{3\times 6}\\
		0_{6\times 3} & \mathbb N
	\end{pmatrix},
\end{equation}
where $h$ stands for the matrix representing the intrinsic metric, $0_{n\times m}$ is a $n\times m$ zero matrix, and $\mathbb N$ is the $6\times 6$ matrix in the lower right corner of~\eqref{eq:HessianExplicit}. The determinant and the inverse of $\mathbb M$ are again easy to compute due to its block diagonal form and one finds
\begin{equation}
	\det\mathbb M = -\frac{\hat c^3}{N^9} \det h\det\mathbb N\quad\text{and}\quad \mathbb M^{-1} = \begin{pmatrix}
		-\frac{N^3}{\hat c} h^{-1} & 0_{3\times 6}\\
		0_{6\times 3} & \mathbb N^{-1}
	\end{pmatrix}.
\end{equation}
Using the expression for $\mathbb M^{-1}$, it follows that $v^\transpose \mathbb M^{-1} v = c^2_{35}\tilde{w}^\transpose \mathbb N^{-1} \tilde{w}$, where $\tilde{w}$ corresponds to the first column of~\eqref{eq:HessianExplicit} divided by $c_{35}$ and without the first four elements. Hence, the determinant of the Hessian can be written as
\begin{equation}\label{eq:IntermediateStep}
	\det H = -\frac{\hat c^3}{N^9}\left(\frac{4}{N^3}\tilde c - c^2_{35} \tilde{w}^\transpose\mathbb N^{-1} \tilde{w} \right)\det h\det\mathbb N.
\end{equation}
Almost all dependence of $\det H$ on the parameters $c_i$ has been separated from its dependence on the field variables. Only $\mathbb N$ still depends on both, the parameters $c_1$ and $c_3$, and some of the field variables. Notice however that $\det \mathbb N$ is a polynomial and that $\mathbb N^{-1}\propto \frac{1}{\det\mathbb N}$ multiplied by $\det\mathbb N$ is also a polynomial in $c_1$ and $c_3$. Hence, it follows that the determinant of the Hessian must have the form
\begin{equation}\label{eq:detHForm}
	\det H = \hat c^3\left(\tilde c\, \text{Poly}(c_1, c_3) - c^2_{35}\, \text{Poly}(c_1, c_3)\right).
\end{equation}
The precise form of the polynomials has to be inferred from computing $\det\mathbb N$ and $\mathbb N^{-1}$ or by a direct computation of the determinant of the original Hessian matrix. Both approaches lead to the final form
\begin{align}\label{eq:DetOfHessian}
	\det H  &= c^5_1\, \hat c^3\left[c_1 \tilde c A - (c^2_{35}-4 c_3 \tilde c) B\right],
\end{align}
which is cubic in $\hat c$, quadratic in $c_{35}$, and linear in $\tilde c$, just as expected from~\eqref{eq:detHForm}, and where $A$ and $B$ are solely functions of the field variables given by\footnote{It appears there is a tension between the factor $\det h$ in~\eqref{eq:IntermediateStep} and the factor $\frac{1}{\det h}$ appearing in the functions $A$ and $B$. However, the latter is due to the fact that both polynomials in~\eqref{eq:detHForm} are proportional to $(\det h^{-1})^2$.}
\begin{align}\label{eq:AandB}
	A &:= \frac{4}{\det h\, N^{18}}h^{\rn{11}}h^{\rn{33}}\left(h^{\rn{11}} h^{\rn{22}}-(h^{\rn{12}})^2\right)\left(h^{\rn{22}}h^{\rn{33}} - (h^{\rn{23}})^2\right)\notag\\
	B &:= \frac{1}{\det h\, N^{18}}\bigg[(h^{\rn{12}})^2 h^{\rn{33}} \left((h^{\rn{13}})^2h^{\rn{22}} - 2 h^{\rn{12}}h^{\rn{13}}h^{\rn{23}} + (h^{\rn{12}})^2 h^{\rn{33}}\right)\notag\\
	&\phantom{:= \frac{1}{\det h\, N^{18}}\bigg[} - h^{\rn{11}}\left((h^{\rn{13}})^2 h^{\rn{22}} - 2 h^{\rn{12}} h^{\rn{13}} h^{\rn{23}} + 4 (h^{\rn{12}})^2 h^{\rn{33}}\right)\left(h^{\rn{22}}h^{\rn{33}}-(h^{\rn{23}})^2\right)\notag\\
	&\phantom{:= \frac{1}{\det h\, N^{18}}\bigg[} (h^{\rn{11}})^2\left((h^{\rn{23}})^2-3h^{\rn{22}}h^{\rn{33}}\right)\left((h^{\rn{23}})^2 - h^{\rn{22}}h^{\rn{33}}\right)\bigg].
\end{align}
Due to the simple structure of~\eqref{eq:DetOfHessian}, it is straightforward to formulate conditions on the parameters $c_i$ which render the rank of $H$ non-maximal:
\begin{align}\label{eq:Conditions}
	\det H = 0\quad &\Longleftrightarrow\quad c^5_1 \hat c^3 = 0\quad\text{or}\quad c_1\tilde c A - \left(c^2_{35}-4 c_3 \tilde c\right) B = 0\notag\\
	&\Longleftrightarrow\quad c_1 = 0\quad\text{or}\quad \hat c = 0\quad\text{or}\quad \tilde c = 0,\, c_{35} = 0\quad\text{or}\quad c_1 = 0,\, c^2_{35} - 4 c_3 \tilde c = 0,\, \tilde c\neq 0.
\end{align}
The second equivalence relation follows from the requirement that the $c_i$ parameters are independent of the field variables, which means that the factors multiplying $A$ and $B$ need to vanish independently. Also, it is necessary to impose $\tilde c \neq 0$ in the last condition because otherwise one simply recovers a subset of the solutions found through the conditions $\tilde c = 0, c_{35} = 0$.

Each one of the four independent parameter conditions~\eqref{eq:Conditions} leads to a different matrix rank for the Hessian and correspondingly to a different number of primary constraints, $M=10-\text{rank} H$. Moreover, each parameter condition defines a hypersurface in the five-parameter space of theories described by $\mathcal Q$. Since these hypersurfaces correspond to theories with different sets of primary constraints, we refer to them as primary sectors. Their properties are summarized in the following table.
\begin{center}
\begin{tabularx}{\textwidth}{|l|>{\centering\arraybackslash}X|>{\centering\arraybackslash}X|>{\centering\arraybackslash}X|>{\centering\arraybackslash}p{4.13cm}|} 
 \hline
 \textbf{Primary sector} & \textbf{I} & \textbf{II} & \textbf{III} & \textbf{IV}\\ \hline
 \textbf{Defining conditions} & $\tilde c = 0$ and $c_{35} = 0$ & $\hat c = 0$ & $c_1 = 0$ & \parbox[c]{\hsize}{$c^2_{35} - 4 c_3 \tilde c = 0$ and $c_1 = 0$$\phantom{I^{I^{I}}_I}$\newline with $\tilde c\neq 0$$\phantom{I^{I}_I}$}\\ \hline
 \textbf{Explicit} & \parbox[c]{\hsize}{$c_4 = -c_1 - c_2 + c_3$\newline $c_5 = -2 c_3$}  & $c_4 = -2c_1 - c_2$ & $c_1 = 0$ & \parbox[c]{\hsize}{$c_1 = 0$ and $c_4 = -c_2 + \frac{c^2_5}{4 c_3}$\newline or\newline $c_1 = 0$, $c_3 =0$, $c_5 = 0$$\phantom{I^{I}_I}$} \\ \hline 
 \textbf{Primary constraints} & $1$ & $3$ & $5$ & $6$ \\ \hline
\end{tabularx}
\end{center}
As we had already seen in the previous section, the conditions $\tilde c = 0, c_{35} = 0$ turn the momentum density conjugate to the lapse function into a primary constraint while $\hat c = 0$ has the effect of eliminating all velocity terms from $\tilde\pi_a$. This independently confirms that these conditions lead to one and three primary constraints, respectively. That $c_1 = 0$ corresponds to five primary constraints would not have been immediately guessed from~\eqref{eq:GeneralMomenta}, but it is clear if we look at the Hessian~\eqref{eq:HessianExplicit}, where $c_1 = 0$ renders the $6\times 6$ submatrix $\mathbb N$ degenerate.

Of course, it is also possible to construct new parameter conditions by identifying where the sectors I--IV intersect each other. However, not all combinations of two sectors lead to new primary constraints: The intersection I~$\cap$~III simply describes a particular subspace of region I and still only gives rise to one primary constraint. Similarly, combining the sectors III and IV does not lead to new constraints because $c_1 = 0$ is already a required condition for sector IV. Also, the sectors I and IV cannot intersect each other because the defining conditions of sector I are incompatible with the requirement $\tilde c\neq 0$ of sector IV. Hence, out of the six possible combinations of two sectors, only three give rise to new primary constraints. These combinations are summarized in the following table.
\begin{center}
\begin{tabularx}{\textwidth}{|l|>{\centering\arraybackslash}X|>{\centering\arraybackslash}X|>{\centering\arraybackslash}X|} 
 \hline
 \textbf{Primary sector} & \textbf{V (GR)} & \textbf{VI} & \textbf{VII} \\ \hline
 \textbf{Intersection of} & I $\cap$ II & II $\cap$ III & II $\cap$ IV \\ \hline
 \textbf{Defining conditions} & $\tilde c = 0$, $\hat c = 0$ and $c_{35} = 0$ & $\hat c = 0$ and $c_1 = 0$ & \parbox[c]{\hsize}{$\hat c = 0$, $c^2_{35}-4c_3 \tilde c = 0$$\phantom{I^{I}_I}$\newline and $c_1 = 0$ with $\tilde c \neq 0\phantom{I^{I}_I}$} \\ \hline
 \textbf{Explicit} & \parbox[c]{\hsize}{$c_3 = -c_1$, $c_4 = -2c_1 - c_2$$\phantom{I^{I}_I}$\newline $c_5 = 2c_1\phantom{I^{I}_I}$} & $c_1 = 0$, $c_4 = -c_2$ & $c_1 = 0$, $c_4 = -c_2$, $c_5 = 0$ \\ \hline 
 \textbf{Primary constraints} & $4$ & $8$ & $9$ \\ \hline
\end{tabularx}
\end{center}
Notice that sector V, characterized by the parameter conditions which turn the momenta conjugate to lapse and shift into primary constraints, has to be the sector which contains GR as a special case. Indeed, one can verify that the GR values~\eqref{eq:GRCoefficients} satisfy the conditions of sector V. Also, observe that due to the independence of the conditions which define the sectors I--IV, the numbers of primary constraints of the sectors V--VII are obtained by simple addition. 

\begin{center}
\begin{figure}[h!]
	\centering
	\includegraphics[width=.5\textwidth]{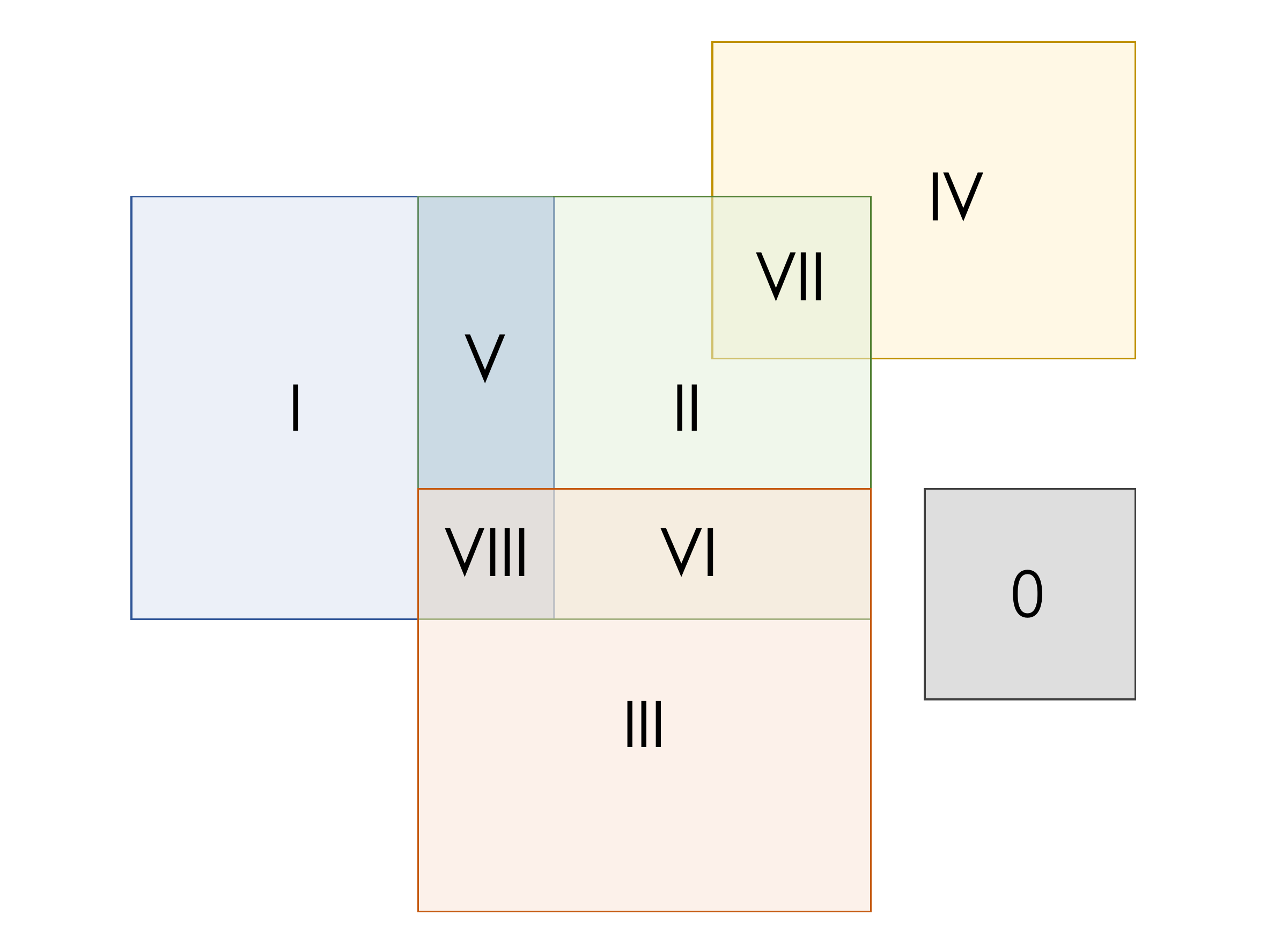}
	\caption{The five-parametric theory space described by the non-metricity scalar $\mathcal Q$ decomposes into nine primary sectors. Four sectors, I--IV, emerge directly from the degeneracy of the Hessian, while the sectors V--VIII describe the intersection of two or three different sectors. The unphysical sector $0$ lies in the complement of I--VIII.}
	\label{fig:Classification}
\end{figure}
\end{center}

Next, we may consider the intersection of three hypersurfaces. A priori there are three possible combinations of the sectors I--IV. However, we have already established above that intersections containing I $\cap$ IV are inconsistent due to the requirement $\tilde c\neq 0$ of sector IV and combining III with IV does not lead to anything new. Hence, the intersections I $\cap$ II $\cap$ IV and II $\cap$ III $\cap$ IV can be excluded and the only remaining option is I $\cap$ II $\cap$ III. The result is summarized in the table below.
\begin{center}
\begin{tabularx}{0.4\textwidth}{|l|>{\centering\arraybackslash}X|} 
 \hline
 \textbf{Primary sector} & \textbf{VIII} \\ \hline
 \textbf{Intersection of} & I $\cap$ II $\cap$ III \\ \hline
 \textbf{Defining conditions} & \parbox[c]{\hsize}{$\tilde c = 0$, $\hat c = 0$,$\phantom{I^{I}_I}$\newline $c_{35} = 0$, $c_1 = 0\phantom{I^{I}_I}$} \\ \hline
 \textbf{Explicit} & \parbox[c]{\hsize}{$c_1 = 0$, $c_3 = 0$,$\phantom{I^{I}_I}$\newline $c_4 = -c_2$, $c_5 = 0\phantom{I^{I}_I}$} \\ \hline 
 \textbf{Primary constraints} & $10$ \\ \hline
\end{tabularx}
\end{center}
With the definition of sector VIII, we have exhausted all possibilities of combining different sectors. The only remaining condition we can impose on the $c_i$ parameters is that they do not satisfy any of the conditions which define the sectors I--VIII. More formally, we can define a new primary sector, which we shall call sector $0$, by the requirement $(c_1, c_2, c_3, c_4, c_5)\in \mathbb R^5\setminus\left(\text{I}\cup\text{II}\cup\cdots\cup\text{VIII}\ \right)$. All nine primary sectors and their interrelations are shown in Figure~\ref{fig:Classification}.
We remark that some of the conditions found in the present context are consistent with prior results obtained through perturbative techniques. In~\cite{Jimenez:2018} it was shown that the perturbative expansion of~\eqref{eq:CGSTG} up to quadratic order around Minkowski space gives rise to an action which in general propagates more than two degrees of freedom. A pure, massless spin-$2$ field only emerges if additional\footnote{In addition to the gauge freedom used to eliminate the connection.} gauge symmetries are stipulated. Requiring invariance under linearized diffeomorphisms led in~\cite{Jimenez:2018} to the parameter conditions $c_3 = -c_1$, $c_4 = -2c_1 - c_2$, and $c_5 = 2c_1$, which we recognize as the explicit conditions which define sector V. Or in other words: These are the conditions which turn the momenta conjugate to lapse and shift into primary constraints. This in turn is consistent with the observation also made in~\cite{Jimenez:2018} in the context of a cosmological mini-superspace model, where it was found that unless $\tilde c = 0$ and $c_{35} = 0$, the lapse becomes a dynamical degree of freedom. 

As discussed in~\cite{Jimenez:2018}, a second gauge symmetry for the perturbative expansion of~\eqref{eq:CGSTG} can be stipulated. The requirement of invariance under transverse diffeomorphisms in conjunction with a Weyl rescaling leads to the parameter conditions $c_3 = -\frac{3}{8} c_1$, $c_4 = -2c_1 - c_2$, and $c_5 = c_1$. These conditions are compatible with sector II, which only requires $c_4 = -2c_1 - c_2$. Indeed, it can be checked by direct computation that these conditions give rise to a Hessian matrix of rank seven, which in turn corresponds to the presence of three primary constraints.  

To conclude this section, we point out that even though a full-fledged Hamiltonian analysis is necessary to reliably count the degrees of freedom and assess the health status of each sector, we can dismiss the sectors $0$, VII, and VIII as unphysical. These sectors propagate ten, at most one, and zero degrees of freedom, respectively. The health status of the sectors I--IV and VI is less clear. In the case of the sectors III, IV, and VI it is not clear what the condition $c_1=0$ actually means in physical terms. However, it should be pointed out that $c_1=0$ removes dynamical terms from the momentum conjugate to the intrinsic metric~\eqref{eq:GeneralMomenta} and that in GR we have $c_1 \neq 0$. This condition therefore seems to be ``subtracting'' from GR, rather than ``adding'' to it, and it seems dubious whether these sectors have anything to do with gravity as we know it.

The primary constraints appearing in the sectors I and II, on the other hand side, have a clear interpretation. However, these constraints are in general not enough to render lapse and shift non-dynamical which would then give rise to ghostly degrees of freedom. A definitive statement can of course only be made after having identified potential secondary constraints.

This finally leaves sector V as the most promising sector. We already know that it contains GR as a special case, but it may also contain viable extensions  and therefore invites further investigations.

\section{Conclusion}
We have performed the first steps of a generalized Hamiltonian analysis of the five-parameter family of non-metricity theories introduced in~\cite{Jimenez:2018} and we have classified all primary constraints in terms of conditions one can impose on the $c_i$ parameters. Our main result is that the theory space can be decomposed into nine sectors, based on the presence or absence of primary constraints, and that certain sectors can be dismissed as unphysical already at this stage.

A key feature of our analysis is the use of the coincident gauge. In this particular gauge the connection vanishes globally and we can use the action~\eqref{eq:CGSTG}, which is purely a functional of the metric, as our starting point. We subsequently performed a $3+1$ decomposition in terms of ADM variables and computed the corresponding conjugate momentum densities. As had to be expected, the momenta conjugate to lapse and shift do not vanish in general, not even for the parameter choices~\eqref{eq:GRCoefficients} which correspond to Coincident General Relativity. However, this is no obstruction since these momenta no longer depend on velocity fields and are therefore turned into primary constraints. 

By systematically examining under which conditions the Hessian matrix~\eqref{eq:HessianExplicit} fails to be invertible, we have identified four independent equations~\eqref{eq:Conditions} which define hypersurfaces in the five-parametric theory space. These hypersurfaces, dubbed primary sectors I through IV, correspond to theories with different sets of primary constraints. Moreover, some of these sectors can intersect each other and thereby give rise to new combinations of primary constraints. In total, there are eight primary sectors I--VIII which are described by certain conditions on the $c_i$ parameters. A ninth sector can be defined by the requirement that the parameters do not obey any of the conditions which define the sectors I--VIII. This sector, dubbed sector $0$, as well as the sectors VII and VIII can be discarded as unphysical because they harbor theories with ten, one, and zero degrees of freedom, respectively. 

The sectors III, IV, and VI all contain the condition $c_1 = 0$ which seems to modify the dynamical behavior of the momentum conjugate to the intrinsic metric and therefore may lead to large deviations from GR which has a $c_1$ parameter different from zero. The primary constraints appearing in the sectors I and II have a clear physical interpretation, but it seems unlikely that they suffice to render lapse and shift non-dynamical and therefore potentially harbor ghostly degrees of freedom. This leaves sector V, which contains GR as a special case, as the most promising sector to look for viable extensions of General Relativity. 

In any case, the next step is to determine the primary Hamiltonian of each sector and compute the Poisson brackets between the constraints and the Hamiltonian. This could potentially lead to secondary constraints and it is a necessary step in order to determine the number of physical degrees of freedom and assess the health status of the various non-metricity theories described by $\mathcal Q$.

\section*{Acknowledgements}
LH is supported by funding from the European Research Council (ERC) under the European Unions Horizon 2020 research and innovation programme grant agreement No 801781 and by the Swiss National Science Foundation grant 179740.


\bibliographystyle{utcaps}
\bibliography{Bibliography}

\end{document}